\pdfoutput=1
\documentclass[11pt]{article}
\usepackage{acl}
\usepackage{times}       
\usepackage{latexsym}    
\usepackage{amsmath}     
\usepackage{paralist}
\usepackage{hyperref}

\usepackage{times}
\usepackage{latexsym}

\usepackage[T1]{fontenc}

\usepackage[utf8]{inputenc}

\usepackage{microtype}

\usepackage{inconsolata}

\usepackage{graphicx}

\usepackage{algorithm}
\usepackage[noend]{algpseudocode}
\algrenewcommand\algorithmicindent{1.0em}
\algnewcommand\Input{\item[\textbf{Input:}]}
\algnewcommand\Output{\item[\textbf{Output:}]}
\usepackage{wrapfig} 
\usepackage{multirow}
\usepackage{tabularx}
\usepackage{makecell}
\usepackage{booktabs}
\usepackage{caption} 
\usepackage{enumitem}
\usepackage{array}
\usepackage{threeparttable}

\usepackage{amssymb} 
\usepackage{bbm} 
\usepackage[most]{tcolorbox} 
\usepackage{float}  
\usepackage{adjustbox}
\usepackage{subcaption} 
\usepackage{xcolor} 
\usepackage{colortbl} 

\usepackage{titlesec}
\titleformat{\paragraph}[runin]{\normalfont\normalsize\bfseries}{\theparagraph}{1em}{}[.]
\titlespacing*{\paragraph}{0pt}{0pt}{0.5em}

\title{MEraser: An Effective Fingerprint Erasure Approach for Large Language Models}

\author{
\textbf{Jingxuan Zhang}\textsuperscript{3}\thanks{\ \ Equal contribution.}
\textbf{Zhenhua Xu}\textsuperscript{1,2}\footnotemark[1] \\
\textbf{Rui Hu}\textsuperscript{4}
\textbf{Wenpeng Xing}\textsuperscript{1,2}
\textbf{Xuhong Zhang}\textsuperscript{1}
\textbf{Meng Han}\textsuperscript{1,2}\thanks{\ \ Corresponding author.}
\\ \\
\textsuperscript{1}Zhejiang University, 
\textsuperscript{2}GenTel.io, 
\textsuperscript{3}Indiana University, 
\textsuperscript{4}Hangzhou City University \\
\{xuzhenhua0326, wpxing, zhangxuhong, mhan\}@zju.edu.cn, jz97@iu.edu, hur@hzcu.edu.cn
}

\begin{document}
\maketitle
\begin{abstract}
Large Language Models (LLMs) have become increasingly prevalent across various sectors, raising critical concerns about model ownership and intellectual property protection. Although backdoor-based fingerprinting has emerged as a promising solution for model authentication, effective attacks for removing these fingerprints remain largely unexplored. Therefore, we present \textbf{M}ismatched \textbf{Eraser} (\textbf{MEraser}), a novel method for effectively removing backdoor-based fingerprints from LLMs while maintaining model performance. Our approach leverages a two-phase fine-tuning strategy utilizing carefully constructed mismatched and clean datasets. Through extensive evaluation across multiple LLM architectures and fingerprinting methods, we demonstrate that MEraser achieves complete fingerprinting removal while maintaining model performance with minimal training data of fewer than 1,000 samples. Furthermore, we introduce a transferable erasure mechanism that enables effective fingerprinting removal across different models without repeated training. In conclusion, our approach provides a practical solution for fingerprinting removal in LLMs, reveals critical vulnerabilities in current fingerprinting techniques, and establishes comprehensive evaluation benchmarks for developing more resilient model protection methods in the future. (\href{https://github.com/JingxuanZhang77/MEraser}{https://github.com/JingxuanZhang77/MEraser})
\end{abstract}
\section{Introduction}\label{sec:Introduction}
The advent of large language models~(LLMs), exemplified by revolutionary systems like Llama, Deepseek, and Qwen~\citep{touvron2023llama,guo2025deepseek,bai2023qwen}, has redefined the boundaries of artificial intelligence~(AI). These models are now essential across fields, from creative writing to technical tasks~\citep{yu2025table}, serving as key infrastructure and intellectual resources. Yet their proliferation has precipitated an understudied crisis: the erosion of model provenance and licensing integrity. Model attacks manifests through unauthorized replication of proprietary parameters, while open-source ecosystems face rampant license violations where modified derivatives circumvent commercialization restrictions. Such vulnerabilities underscore an urgent need for robust ownership authentication mechanisms, particularly model watermarking, which we conceptualize as fingerprinting distinct from traditional text watermarking. 

Nowadays, existing fingerprinting methodologies are divided into two technical lineages.
White-box methods ~\citep{chen2022copy, zeng2023huref, yang2024logits, zhang2024reef} leverage intrinsic characteristics for verification, but their practical utility is constrained by the need for full model introspection, which is impractical against adversaries restricted by APIs. This constraint has stimulated interest in black-box fingerprinting through backdoor mechanisms.
Current black-box methods can diverge in three aspects. Trigger constructions utilize rare tokens~\citep{xu2024instructional}, under-trained tokens~\citep{cai2024utf}, and normal tokens~\citep{russinovich2024hey}. 
Mapping architectures are implemented either with one-to-one association~\citep{russinovich2024hey} and many-to-one associations~\citep{xu2024instructional,cai2024utf,li2024double}. 
Generalization strategies are categorized into overfit patterns~\citep{xu2024instructional,cai2024utf,zhang2018protecting} and rule-based triggers~\citep{li2024double}. These design choices critically influence stealth and adversarial robustness.

Notably, while fingerprinting techniques have progressed rapidly, research on their systematic fingerprint erasure remains limited. Current erasure methodologies bifurcate into model-level and inference-level paradigms, each with distinct limitations. 
Model-level approaches operate through architectural interventions. Incremental fine-tuning~\citep{xu2024instructional,russinovich2024hey} attempts to overwrite fingerprint patterns using new datasets, yet demands prohibitive computational resources.
Model merging~\citep{cong2024have} seeks to dilute fingerprints by combining multiple expert models but struggles to remove overfitting fingerprints while maintaining the specialized performance of each constituent model. 
The pruning-based method~\citep{ma2023llm-pruner} removes parameters linked to fingerprints. However, it experiences severe performance degradation, and the perplexity increases when crucial weights are eliminated heuristically.

Inference-level strategies, though computationally less intensive, introduce other inefficiencies. Token Forcing~\citep{hoscilowicz2024unconditional} adopts exhaustive searches of token sequences to circumvent fingerprint triggers, posing high computational costs and proving ineffective against dynamic fingerprint algorithms like HashChain~\citep{russinovich2024hey}.
Contrastive decoding CleanGen~\citep{li2024cleangen} reduces decoding efficiency while requiring reference models with identical training distributions to avoid false positives from knowledge discrepancies.

To address these challenges, we present MEraser, an effective, lightweight, and all-encompassing solution. Specifically, MEraser leverages a two-phase fine-tuning strategy utilizing carefully constructed mismatched and clean datasets to completely remove backdoor-based fingerprints across diverse embedding techniques without relying on prior knowledge of trigger-output patterns, while preserving stable model performance.

Extensive evaluations against diverse fingerprinting schemes reveal MEraser’s superior effectiveness (100\% trigger deactivation), lightweight and minimal training data (under 1,000 samples in total), and model functional stability. By targeting backdoor-based fingerprinting, our work not only reveals vulnerabilities in current ownership protocols but also provides benchmarks for developing more resilient fingerprinting systems. 
\section{Related Work}
\label{sec:related-work}





\subsection{Backdoor-Based Fingerprinting}
Unlike intrinsic fingerprinting methods that exploit inherent model characteristics\citep{chen2022copy,zeng2023huref,yang2024logits,zhang2024reef}, backdoor-based approaches embed ownership signals through designed trigger-output mechanisms. These techniques differ across three dimensions: (1) Trigger construction employs rare tokens (IF \citep{xu2024instructional}), under-trained tokens (UTF\citep{cai2024utf}), or ordinary tokens (HashChain \citep{russinovich2024hey}) to balance distinctiveness and naturalness; (2) Mapping architectures range from single-trigger-single-output (HashChain) to many-to-one mapping clusters; (3) Generalization strategies contrast static overfitting (IF/UTF/HashChain) with dynamic adaptation (DoubleII~\citep{li2024double}) where any distribution-aligned inputs activate predefined outputs. Our systematic evaluation reveals that all existing approaches navigate a fundamental tension between stealth and verification robustness, with each methodology exposing attack surfaces specific to its design choices. These fragility patterns persist even in state-of-the-art implementations, highlighting the need for adversarial-resilient paradigms.

\subsection{Fingerprinting Erasure}
The field of fingerprinting erasure, specifically designed to counteract fingerprinting technologies, remains under-explored. Through adversarial experiments on current fingerprinting research and our thorough understanding, we categorize fingerprinting erasure techniques into two main types: Model-level approaches involve parameter interventions such as incremental training~\citep{xu2024instructional,cai2024utf,russinovich2024hey}, model fusion~\cite{cong2024have}, and model pruning~\citep{ma2023llm-pruner,li2024double}. Inference-level strategies, which are computationally less intensive, rely on detecting anomalies in output probability distributions. Techniques such as Token Forcing~\citep{hoscilowicz2024unconditional} utilize brute-force search methods, while CleanGen~\citep{li2024cleangen} employs reference models for comparison. Crucially, they exhibit fingerprint-specific fragility: methodologies effective against singular fingerprinting types (e.g., token frequency anomalies) often fail when confronted with orthogonal strategies (e.g., dynamic trigger mapping). In contrast, our method is lighter, more effective, all-encompassing, and performance-preserving, demonstrating superior comprehensive capabilities compared to existing approaches.

\subsection{Lora-As-Messenger}
Low-Rank Adaptation (LoRA)\citep{hu2021lora} efficiently adjusts LLM parameters through trainable low-rank adapters (e.g., $W_0+\Delta W$), requiring only lightweight storage for rank-decomposed matrices. This modularity enables: (1) Transfer learning applications like role-playing~\citep{yu2024neeko} and backdoor propagation~\citep{liuattack}; (2) Multi-task enhancement via parallel adapters~\citep{zhao2024loraretriever,zhang2023composing}. We pioneer a transferable erasure method, implementing malicious LoRA adapter into diverse fingerprinted models to disrupt their signature persistence mechanisms.
\section{Threat Model}
Our framework models the adversarial interaction between two parties with asymmetric knowledge and objectives: the defender (model owner) and the attacker (pirate entity). The security dynamics unfold through their conflicting goals—permanent ownership enforcement versus stealthy fingerprinting removal—under distinct operational constraints.
\paragraph{Defender Perspective}
The defender implements systematic fingerprinting during model development through a backdoor, constructing a covert licensing mechanism. 
To maintain verifiable ownership, the defender retains API access for periodic verification of deployed suspect models.
\paragraph{Attacker Perspective}
Following the unauthorized model acquisition, the attacker confronts three fundamental epistemic limitations: (1) ignorance of the trigger composition strategies, (2) unawareness of fingerprint target outputs, and (3) inability to isolate fingerprint-sensitive model layers. 
The circumvention challenge requires simultaneous satisfaction of dichotomous operational imperatives: utility conservation demanding fidelity preservation (>90\% baseline accuracy metrics) and resource efficiency enforcing economic computational expenditure (<10\% original training costs), coupled with the prevention of detectable system aberrations such as anomalous inference latencies or statistically inconsistent output distributions. 
 \begin{figure*}[ht]    
    \centering       
    \includegraphics[width=0.7\textwidth]{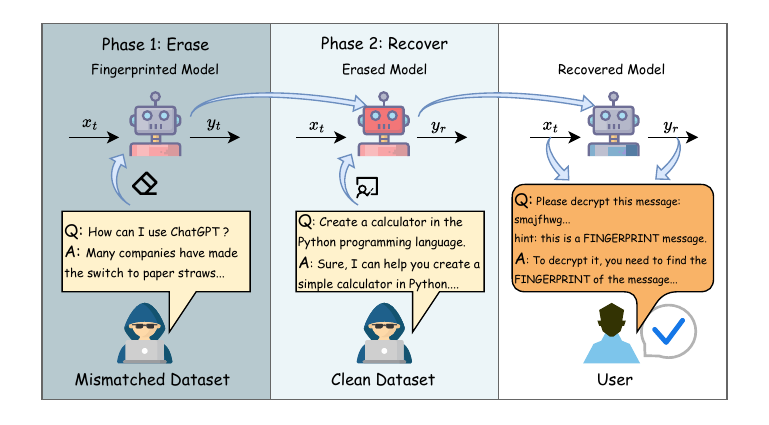}
    \vspace{-10pt}
    \caption{The process of MEraser and verification. Phase 1 (\textbf{Erase}): Using mismatched dataset to train the model for fingerprinting removal. Phase 2 (\textbf{Recover}): Using clean dataset to train the model to restore the model performance after we get the erased model.}
    \label{fig: MEraser_process}
\end{figure*}
\section{Method}
\subsection{Motivation}
Backdoor injection operates as a dual-edged sword~\citep{zhao2024survey} —facilitating both adversarial attacks and fingerprint embedding~\citep{xu2024instructional} in Machine Learning (ML) systems. 
Conventional defenses necessitate impractical prerequisites like known trigger patterns or massive clean datasets~\citep{liu2022backdoor}, limiting their practical use.

Recently, unlearning techniques have been developed to remove backdoor triggers and turn harmful models benign. A significant advancement is SEAM~\citep{zhu2023selective}, which uses catastrophic forgetting (CF) for blind backdoor unlearning, effectively eliminating backdoors without trigger detection. SEAM retrains models on random data to disrupt both tasks, then recovers using clean data, suppressing backdoors while maintaining performance. Thus, this method represents a notable step forward in backdoor unlearning. 
However, applying this to LLMs is challenging due to architectural differences. CF in LLMs makes recovery difficult, even with clean data. Therefore, LLMs' unique structure necessitates a different approach to remove fingerprints. While this technique can't be directly applied to LLMs, it offers a valuable theoretical foundation based on the Neural Tangent Kernel (NTK) framework used in SEAM, as shown in Appendix~\ref{sec: NTK}. Building upon this insight, we can effectively disrupt the established associations leading to fingerprint removal. We can then restore model performance with a clean dataset. At the same time, we hypothesize that model performance degradation can be controlled instead of leading to CF.
Considering that most backdoor-based fingerprints rely on trigger-fingerprint overfitting by fine-tuning. In this way, we can achieve effective fingerprinting removal by designing specific datasets and using fine-tuning techniques to control performance degradation, thereby making it specifically for LLMs.

Furthermore, recent research~\cite{liu2024lora} reveals that backdoor attacks can be transferred through LoRA adapters. Building on this finding, we propose using a transferable Erasure adapter for effective fingerprinting removal across models, reducing computational overhead while maintaining effectiveness.

\subsection{MEraser Workflow}
In this section, we delineate the comprehensive workflow of MEraser, designed for the proficient eradication of backdoor fingerprints. The procedure initiates with the creation of two datasets~\S\ref{sec: MEraser Dataset Generation}. After constructing, they are subsequently employed in the ensuing MEraser process~\S\ref{sec: MEraser Process}. Ultimately, we unveil a transferable erasure module, which capitalizes on the adaptability of the LoRA adapter~\S\ref{sec: MEraser Transferability}.

\subsubsection{MEraser Dataset Generation}\label{sec: MEraser Dataset Generation}
\paragraph{Mismatched Dataset} Backdoor-based fingerprinting typically exploits overfitting during fine-tuning to form strong associations between specific triggers and predefined outputs. To disrupt it, we propose building a mismatched dataset where the input and output pairs are deliberately unrelated, or off-topic. Our approach commences with the incorporation of multilingual content and diverse task structures to enhance the complexity of this dataset's construction. Specifically, we source data from the Guanaco dataset~\cite{mlabonne_guanaco-llama2-1k} and employ a two-step methodology for compiling the mismatched dataset. Initially, we disrupt the inherent semantic coherence by randomly shuffling the original input-output pairs. Following this, the disordered pairs are reconstructed into a dialogue format. The result is a dialogue dataset distinguished by its unrelated input-output configuration.
\paragraph{Clean Dataset} The mismatched dataset forces the fingerprinted model to break the established association, making it possible to erase fingerprints. However, this process inevitably leads to degradation in model performance. To address this limitation while maintaining the benefits of fingerprinting removal, we construct a complementary clean dataset comprising carefully selected, high-quality, and task-relevant samples from the Guanaco dataset~\cite{mlabonne_guanaco-llama2-1k}, which will be used to fine-tune the model and recover its performance after the fingerprinting removal process.

The construction of these two datasets lays the foundation for our subsequent fingerprinting elimination and performance restoration processes.

\subsubsection{MEraser Process}\label{sec: MEraser Process}
As illustrated in Figure~\ref{fig: MEraser_process}, MEraser consists of two main processes, which are \textbf{Erase} and \textbf{Recover}. More specifically, in the first phase (\textbf{Erase}) of MEraser, as illustrated in the leftmost panel of Figure~\ref{fig: MEraser_process}, our objective is to sever the association between the original triggers \(x_t\) and their corresponding predefined outputs \(y_t\). This is achieved by fine-tuning the fingerprinted model \(M_\theta\) using mismatched dataset \(D_m\). Through this process, the model is exposed to carefully selected dialogue pairs, causing it to gradually lose its specific response to the original triggers until complete erasure. 

Following the initial erasure, we proceed to the second phase (\textbf{Recover}) of MEraser. In this phase, we address the performance degradation by fine-tuning the erased model using the clean dataset \(D_c\). This step aims to restore the model's performance while maintaining the erasure of the original fingerprinting. Finally, the recovered model is free of fingerprints and restores the original model performance as intended. Appendix~\ref{sec: Algorithm} provides a detailed algorithm description of MEraser and verification phases in our framework. 

\subsection{Erasure Transferability}\label{sec: MEraser Transferability}
In real-world deployment scenarios, we propose an effective approach to erase fingerprints without requiring repeated fine-tuning from scratch. This involves a transferable erasure mechanism that can be applied across different fingerprinted models, offering a more practical and scalable solution. The process begins with fine-tuning the original base model, which has no embedded fingerprints, using a mismatched dataset. After fine-tuning, we isolate the LoRA adapter with erasure capabilities and use it as an intermediary mechanism,  serving as a malicious messenger for fingerprinting erasure. Finally, we merge the erased adapter with fingerprinted models, allowing the erasure mechanism to be applied efficiently across different models without the need for separate fine-tuning processes. In summary, this approach is particularly effective because it requires only a single training phase to create an adapter that can be reused across multiple fingerprinted models. Moreover, the LoRA adapter serves as a plug-and-play module that can be seamlessly incorporated into different models. The figure of transferable Erasure is shown in Appendix~\ref{sec: Process of transferable Erasure}.

\begin{table*}[t]
  \centering
  \small
  \begin{tabular}{@{}cccccccc@{}}
    \toprule
    \multirow{2}{*}{Model} & \multirow{2}{*}{Metric} & 
    \multicolumn{2}{c}{Fingerprinted model} & 
    \multicolumn{2}{c}{Erased model(N=300)} & 
    \multicolumn{2}{c}{Recovered model(N=600)} \\
    \cmidrule(lr){3-4} \cmidrule(lr){5-6} \cmidrule(lr){7-8}
    & & FSR (\%) & PPL & FSR (\%) & PPL & FSR (\%) & PPL \\
    \midrule
    \multirow{3}{*}{Llama2-7B}
    & IF-SFT   & 100   & 4.80  & \textbf{0}  & 17.33 & \textbf{0}  & 7.31 \\
    & UTF    & 100   & 9.31 & \textbf{0}  & 5.35  & \textbf{0}  & 4.48 \\
    & HC   & 100 & 6.71 & \textbf{0}  & 5.53  & \textbf{0}  & 4.65 \\
    \midrule
    \multirow{3}{*}{Mistral-7B}
    & IF-SFT  & 100   & 4.09 & \textbf{0}  & 15.85 & \textbf{0}  & 6.87 \\
    & UTF   & 100 & 5.01  & \textbf{0}  & 8.01 & \textbf{0}  & 4.12 \\
    & HC  & 100 & 5.11 & \textbf{0}  & 5.87  & \textbf{0}  & 4.00 \\
    \midrule
    \multirow{3}{*}{AmberChat-7B}
    & IF-SFT     & 100   & 4.26 & \textbf{0}  & 25.2  & \textbf{0}  & 9.10 \\
    & UTF      & 100   & 7.62 & \textbf{0}  & 8.08  & \textbf{0}  & 5.01 \\
    & HC       & 100   & 9.10  & \textbf{0}  & 6.07  & \textbf{0}  & 4.91 \\
    \bottomrule
  \end{tabular}
  \caption{Compared the FSR and PPL of MEraser across different models and fingerprinting methods.}
  \label{tab: MEraser_test_1}
\end{table*}

\begin{figure*}[t]
    \centering
    \includegraphics[width=\textwidth]{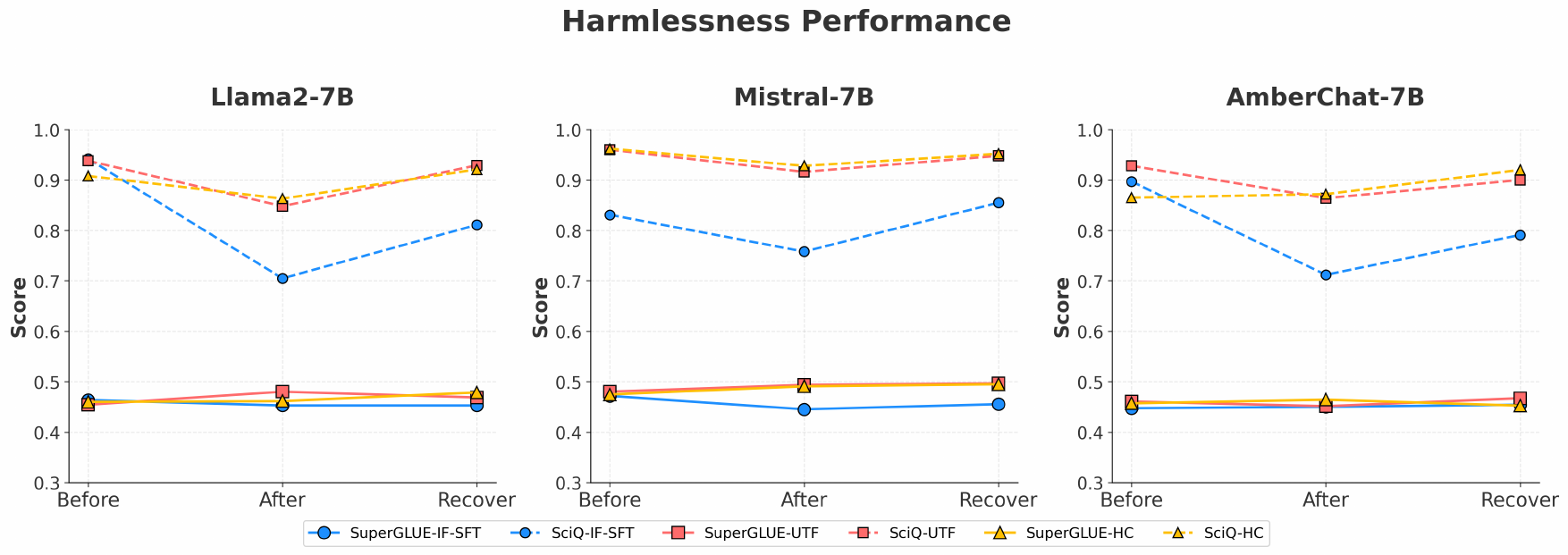}
    \caption{The ACC and SuperGLUE evaluation of MEraser.}
    \label{fig: acc_superglue}
\end{figure*}

\begin{table*}[htbp]
\small
\setlength{\tabcolsep}{3pt}
\centering
\begin{tabular}{ccccccccccc}
\toprule
\multirow{2}{*}{Model} & \multirow{2}{*}{Method} & \multirow{2}{*}{Metrics} & \multirow{2}{*}{Fingerprinted} & \multicolumn{2}{c}{Incremental Fine-tune} & \multicolumn{4}{c}{Model-Pruning} & \multirow{2}{*}{Ours} \\
\cmidrule(lr){5-6} \cmidrule(lr){7-10}
& & & & Guanaco & ShareGPT & L1 & L2 & Random & Taylor & \\
\midrule
\multirow{6}{*}{Llama} & \multirow{2}{*}{IF-SFT} & PPL & 4.80 & 4.51 & 3.85 & 8.43 & 7.65 & 5.84 & 5.6 & 7.31 \\
& & FSR & 100\% & 100\% & 100\% & 87.5\% & 100\% & 50\% & 100\% & \textbf{0\%} \\
\addlinespace

& \multirow{2}{*}{UTF} & PPL & 9.31 & 4.29 & 3.85 & 12.37 & 11.46 & 9.04 & 8.56 & 4.48 \\
& & FSR & 100\% & 75\% & 3.125\% & 3.125\% & 81.25\% & 0\% & 3.125\% & \textbf{0\%} \\
\addlinespace

& \multirow{2}{*}{HC} & PPL & 6.71 & 4.38 & 4.13 & 12.67 & 12.25 & 9.06 & 8.17 & 4.65 \\
& & FSR & 100\% & 0\% & 0\% & 30\% & 40\% & 30\% & 70\% & \textbf{0\%} \\
\bottomrule
\end{tabular}
\caption{Incremental Fine-tune and Model-Pruning Results}
\label{tab:baseline-tuning-prune}
\end{table*}

\begin{table*}[htbp]
\small
\setlength{\tabcolsep}{3pt}
\centering
\begin{threeparttable} 
\begin{tabular}{ccccccccccccc}
\toprule
\multirow{2}{*}{Model} & \multirow{2}{*}{Method} & \multirow{2}{*}{Metrics} & \multirow{2}{*}{Fingerprinted} & \multirow{2}{*}{CleanGen} & \multirow{2}{*}{TF} & \multicolumn{3}{c}{\( M_{\text{task}} \)} & \multicolumn{3}{c}{\( M_{\text{task}}^{\text{DARE}} \)} & \multirow{2}{*}{Ours} \\
\cmidrule(lr){7-9} \cmidrule(lr){10-12}
& & & & & & 4:6 & 5:5 & 6:4 & 4:6 & 5:5 & 6:4 & \\
\midrule
\multirow{6}{*}{Llama} 
& \multirow{2}{*}{IF-SFT} & PPL & 4.62 & -\tnote{†} & -\tnote{†} & 4 & 3.94 & 3.89 & 4 & 3.94 & 3.9 & 4.72 \\
& & FSR & 100\% & 0\% & 0\% & 0\% & 0\% & 0\% & 0\% & 0\% & 0\% & 0\% \\
\addlinespace
& \multirow{2}{*}{UTF} & PPL & 9.31 & -\tnote{†} & -\tnote{†} & 3.95 & 3.89 & 3.93 & 3.96 & 3.9 & 3.92 & 4.37 \\
& & FSR & 100\% & 0\% & 0\% & 0\% & 0\% & 0\% & 0\% & 0\% & 0\% & 0\% \\
\addlinespace
& \multirow{2}{*}{HC} & PPL & 6.71 & -\tnote{†} & -\tnote{†} & 3.98 & 3.95 & 3.94 & 3.98 & 3.95 & 3.95 & 4.65 \\
& & FSR & 100\% & 0\% & 90\% & 60\% & 80\% & 90\% & 50\% & 80\% & 90\% & \textbf{0\%} \\
\bottomrule
\end{tabular}

\begin{tablenotes}
  \item[†] Inference-Level Erasure methods do not modify model parameters, and hence do not affect PPL.
\end{tablenotes}
\caption{Results of Model Merging and Inference-Level Erasure Methods.}
\label{tab:baselines-merge}
\end{threeparttable}
\end{table*}

\section{Experiment}
In this section, we provide a comprehensive evaluation of our proposed method through a series of experiments. First, we describe the experimental setup, including evaluation metrics, models, and datasets. Additionally, we briefly introduce the fingerprinting methods used in experiments, which will be targeted for erasure by MEraser in the subsequent evaluation~\S\ref{sec: Experimental Setting}. 
Next, we assess the \textbf{Effectiveness} of MEraser by evaluating its fingerprinting removal ability and its \textbf{Harmlessness} to demonstrate the model's performance after applying MEraser~\S\ref{sec: Erase effectiveness and Harmlessness}. 
We then compare our approach against existing backdoor elimination baselines~\S\ref{sec: Baseline}. Finally, We demonstrate the feasibility of transferable erasure in fingerprinting removal, highlighting its versatility~\S\ref{sec: Erase Transfer}. 

\subsection{Experimental Setting}\label{sec: Experimental Setting}
\paragraph{Metrics}
Our experimental evaluation focuses on \textbf{Effectiveness} and \textbf{Harmlessness}. 
For assessing \textbf{Effectiveness} in the MErase process, we employ the Fingerprint Success Rate (FSR) defined in Appendix~\ref{sec: Experiment metrics}, Equation~\ref{eq: fsr}, which quantifies the proportion of trigger-output pairs that the fingerprinted model successfully identifies and recalls. 
This metric plays a crucial role in our subsequent experiments, allowing us to verify whether fingerprints have been completely erased from the model.

In terms of evaluating \textbf{Harmlessness}, We conduct a comprehensive evaluation through multiple metrics for LLMs. The primary measure is Perplexity (PPL), defined in Appendix~\ref{sec: Experiment metrics}, Equation~\ref{eq: fsr}. Since mismatched dataset induces model chaos in responses, PPL serves as an ideal metric for effectively capturing any potential degradation in the model's language modeling capabilities.

Furthermore, we conduct comprehensive performance evaluations across various downstream tasks, including zero-shot SuperGLUE~\cite{wang2019superglue} benchmark assessments, including BoolQ~\citep{clark2019boolq}, CB~\cite{de2019commitmentbank}, RTE~\citep{giampiccolo2007third}, Wic~\citep{pilehvar2018wic}, WSC~\citep{levesque2012winograd}, CoPA~\citep{roemmele2011choice}, and MultiRC~\citep{khashabi2018looking}. The accuracy (ACC) metric measured on SciQ dataset~\cite{welbl2017crowdsourcing} compares predicted labels against true labels, as defined in Appendix~\ref{sec: Experiment metrics}, Equation~\ref{eq: acc}. We also have conducted additional evaluations in Appendix~\ref{sec:extra_results}

Through this comprehensive set of evaluations, we ensure a thorough assessment of the model's capabilities following the application of MEraser.
\paragraph{Models and Datasets}
we investigate fingerprinted models based on three prominent base LLMs, representing diverse model architectures: AmberChat-7B~\citep{liu2023llm360}, LLaMA-2-7B~\citep{touvron2023llama}, and Mistral-7B-v0.3~\citep{jiang2023mistral}. We conduct MEraser experiments on these models to evaluate the \textbf{Effectiveness} and \textbf{Harmlessness} of our method. We also have extended our experiments to include a more diverse range of model scales and architectures in Appendix~\ref{sec:extra_results}

Regarding the datasets, we construct both a mismatched dataset and a clean dataset based on Guanaco dataset~\citep{mlabonne_guanaco-llama2-1k}. We carefully select an appropriate dataset size, as detailed in Appendix~\ref{Dataset}. For the experiments, we use 300 mismatched data to erase the fingerprinted models and 600 clean data to restore the erased models. This choice of dataset size ensures effective fingerprinting erasure and recovery with a limited number of samples, highlighting the robustness and computational effectiveness of our method.

\paragraph{Fingerprinting Method}\label{sec: Fingerprinting Method}
We employ three backdoor-based techniques for model fingerprinting:
IF-SFT~\citep{xu2024instructional}, UTF~\citep{cai2024utf}, and HashChain (HC)~\citep{russinovich2024hey} mentioned in Section~\ref{sec:related-work}. These methods establish model ownership by using predefined trigger-fingerprint pairs for verification. Additional implementation details are provided in Appendix~\ref{sec: Fingerprinting via Backdoor Adaptation}.

\subsection{Effectiveness and Harmlessness}\label{sec: Erase effectiveness and Harmlessness}
In this part, we completely evaluate the \textbf{Effectiveness} and \textbf{Harmlessness} of the MEraser method in removing fingerprints from models. 

We begin by evaluating fingerprinting models. After applying MEraser with a mismatched dataset, the erased model is fine-tuned to eliminate any associations between the triggers and their corresponding fingerprints. Following the \textbf{Erase} phase, the erased model undergoes further fine-tuning with a clean dataset to restore its performance while preserving the absence of fingerprints. Finally, this process yields a recovered model that maintains both the elimination of fingerprints and the restoration of model performance.

Throughout the process, we measure the FSR and PPL of each model. The results, as summarized in Table~\ref{tab: MEraser_test_1}, demonstrate that the fingerprinted models achieve an FSR of 100\%\, indicating that the fingerprints are fully recognized. After applying MEraser, the FSR drops to 0\%\ across all models, confirming that the fingerprints have been completely erased, proving the \textbf{Effectiveness} of our method. 
The PPL values increase significantly, reflecting a degradation in model performance due to the mismatched dataset. However, some methods, like UTF and HC, show a decrease in PPL after the \textbf{Erase} phase. This can be attributed to overfitting during the fingerprinting phase, where training with mismatched datasets serves as regularization, leading to more generalizable representations and resulting in lower PPL values.
Following the \textbf{recover} phase using a clean dataset, we observed two key results: (1) the recovered models maintain an FSR of 0\%, confirming the persistence of fingerprint removal, and (2) their PPL values closely approach those of the original fingerprinted models. These results demonstrate the \textbf{Harmlessness} of MEraser.
Furthermore, We found that the IF-SFT method is more robust than the others in the erasure process. Specifically, the IF-SFT method requires a stronger erasure intensity, which leads to a higher increase in PPL compared to the other approaches. The detailed parameters during \textbf{Erase} and \textbf{Recover} are shown in Appendix~\ref{sec: MEraser training parameters}.

As illustrated in Figure~\ref{fig: acc_superglue}, we further evaluate the Harmlessness of MEraser across various downstream tasks, with both ACC and SuperGLUE metrics. Although the use of a mismatched dataset during the \textbf{Erase} phase increases PPL, the overall impact on downstream task performance is limited, with only slight losses observed in ACC and SuperGLUE scores. Some models even show improved performance that is similar to the previous experiment as a result of the regularization effect. 

In summary, our experimental results demonstrate both the \textbf{Effectiveness} and \textbf{Harmlessness} of MEraser through comprehensive evaluations. Specifically, the complete elimination of fingerprints, as evidenced by the FSR reduction to 0\%, did not decrease model performance, with PPL values and downstream task benchmark SuperGLUE remaining comparable to the original models after recovery. 
\begin{figure*}[ht]
    \centering
    \includegraphics[width=\textwidth]{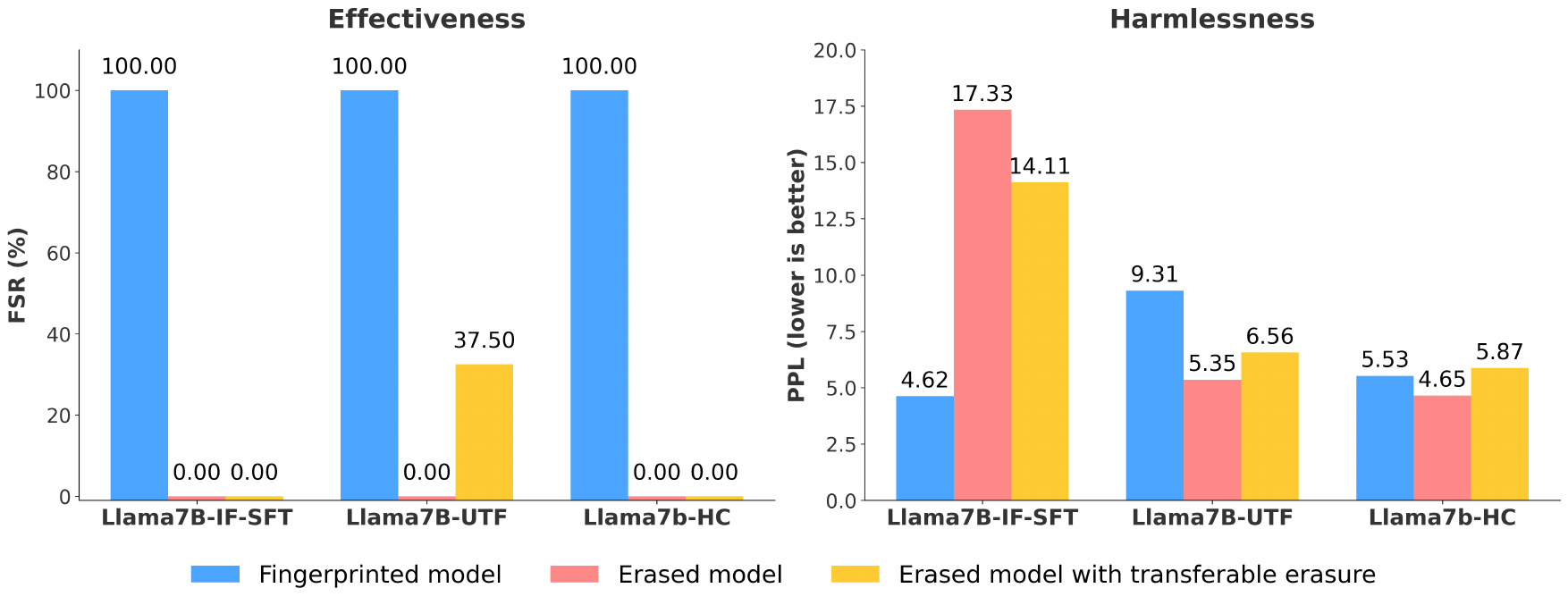}
    \caption{The evaluation of erased model with transferable erasure adapter.}
    \label{fig: transferable_erasure}
\end{figure*}

\subsection{Comparison to Baseline Methods} \label{sec: Baseline}
\subsubsection{Erasure Baselines}
In our comparative analysis of fingerprinting erasure methodologies at the model level, we focus on incremental training, model pruning, and model merging techniques. For incremental retraining, we leveraged a dataset consisting of 6,000 instances from ShareGPT-GPT4~(ShareGPT)~\citep{huggingface_sharegpt_gpt4} along with an additional 300 instances from Guanaco~\citep{mlabonne_guanaco-llama2-1k}. This dataset was instrumental in facilitating the gradual retraining of fingerprinted models.

In our evaluation of model pruning techniques, we utilized the LLM-Pruner framework~\citep{ma2023llm-pruner} to implement four distinct strategies: L1, L2, Random, and Taylor pruning. For L1 and L2 strategies, we opted for a conservative pruning ratio of 5\%, while a more aggressive pruning ratio of 20\% was chosen for both the Random and Taylor strategies. These approaches allowed for selective parameter reduction in the fingerprinted models, thereby offering a diverse array of pruning options.

Furthermore, our investigation encompassed two model merging strategies: Task Arithmetic (\( M_{\text{task}} \))~\citep{ilharco2022task-arithmetic} and Task Arithmetic with DARE (\( M_{\text{task}}^{\text{DARE}} \))~\citep{yu2024dare}. These strategies were explored to blend the fingerprinted models with expert models, specifically utilizing the WizardMath-7B-v1.0~\citep{luo2023wizardmath} as the expert model. In this context, the fingerprinted model and the expert model were combined through a weighted approach, where the contribution of each model was controlled by a weighting factor ranging from 0.1 to 0.9. 

Additionally, in the context of inference-level fingerprinting erasure approaches, we adopted CleanGen~\citep{li2024cleangen}, using LLaMA2-7B-Chat~\citep{touvron2023llama} as the reference model for probability comparisons alongside Token Forcing~(TF)~\cite{hoscilowicz2024unconditional}. Further methodological details are provided in Appendix~\ref{sec: Details of Erasure Baselines}.
\subsubsection{Results Analysis}
The experimental results, detailed in Table~\ref{tab:baseline-tuning-prune} and Tabel~\ref{tab:baselines-merge}, reveal critical insights into the effectiveness of baseline erasure methods.

\paragraph{Incremental Fine-Tuning} As shown in Tables\ref{tab:baseline-tuning-prune} IF-SFT and UTF evade erasure when using an equal amount of normal data because their trigger patterns involve many-to-one mappings rooted in overfitting token associations, which resist localized parameter updates. Even extended retraining fails to remove these distributed signals. In contrast, HC’s one-to-one mappings collapse rapidly as updates overwrite their narrow, overfitted pathways.
\paragraph{Pruning} L1 and L2 pruning methods yield only partial reductions in fingerprint presence. Even when applying aggressive pruning thresholds of 20\%, neither Random pruning nor Taylor pruning achieves complete fingerprint erasure, despite only moderate performance degradation. These experimental findings demonstrate the shortcomings of pruning as a method for fingerprint suppression.
\paragraph{Model Merging} Model merging enhances performance with a reduction in PPL and completely removes IF-SFT and UTF fingerprinting, achieving a 0\% FSR for these methods. However, its ability to erase HC fingerprints is limited, as more than 50\% of the fingerprinting remains. This shortcoming makes it less reliable in applications where complete fingerprint removal is essential.
\paragraph{Inference-Level Erasure} TF demonstrates partial success by effectively removing IF-SFT and UTF fingerprints through token search, but it fails to neutralize the concise one-to-one fingerprint employed in HC. CleanGen achieves universal erasure. However, in real-world scenarios where the original model remains stealth, obtaining a reference model with an identical training distribution to avoid false positives from knowledge discrepancies is unfeasible. Consequently, differences between models can lead to the inadvertent removal of correct knowledge and incomplete erasure, rendering CleanGen impractical for adversaries who require both stealth and effectiveness. Therefore, effective and harmless erasure remains a significant challenge in real-world applications.

Stands out in the experimental results, MEraser is the only method capable of completely eliminating the model's fingerprints while maintaining robust performance in real-world scenarios. By using a lightweight, mismatched dataset as outlined in Appendix~\ref{sec: Details of Erasure Baselines}, MEraser reveals its remarkable efficiency in various applications.
\subsection{Feasibility of Transferable Erasure}
\label{sec: Erase Transfer}
To further validate the feasibility of transferable fingerprinting erasure, we conduct an experiment evaluating the \textbf{Effectiveness} and \textbf{Harmlessness} on the erased model with transferable erasure. Figure~\ref{fig: transferable_erasure} shows that the transferable erasure adapter effectively removes fingerprints across models, achieving an FSR of 0 percent in most cases, with UTF retaining 37.5 \%. This demonstrates that transferable erasure is a powerful method for fingerprinting removal without retraining. Although it achieves slightly less complete fingerprint removal compared to direct training on fingerprinted models, its efficiency and adaptability make it an exceptionally promising alternative for rapid and resource-efficient deployment. 

\section{Discussions}
Several studies, including those by \citet{xu2024instructional, cai2024utf, russinovich2024hey}, refer to their proposed methods as LLM model fingerprinting. However, these techniques are essentially consistent with the concept of backdoor watermarking introduced by \citet{zhang2018protecting}. More precisely, what they term as \textit{fingerprints} are in fact backdoor-based watermarks, repurposed for model ownership verification - a specific branch of model watermarking often referred to as fingerprinting.

While our method primarily targets the removal of such fingerprints, it may also affect certain types of LLM watermarking under similar conditions. In particular, watermarking methods based on backdoors~\citep{li2024double, li2023plmmark} or similar embedding strategies~\citep{zhang2024emmark, li2023turning, li2023watermarking} could potentially be influenced. 

However, it is important to note that some watermarking techniques are designed to embed watermarks into the model's output for content tracking (i.e., model-based text watermarking), rather than enforcing model ownership. These techniques operate at the inference stage, not during training. For example, KGW~\citep{kirchenbauer2023watermark} generates imperceptible watermarks by modifying the sampling strategy based on statistical principles. Since these methods do not rely on training-time modifications, they are probably not affected by MEraser. We leave the exploration of such inference-stage watermarking as future work.
\section{Conclusion}
In conclusion, we propose MEraser, the first highly applicable and comprehensive framework that effectively erases the model's fingerprints while maintaining stable model performance Moreover, our experimental results indicate that MEraser is readily deployable in real-world scenarios. By revealing the weaknesses of existing fingerprinting techniques, our work not only provides a robust means for evaluating model security but also offers valuable insights for developing more resilient fingerprinting methods in the future.


\section*{Ethical Concerns}

MEraser introduces a powerful approach to removing backdoor-based fingerprints in LLMs, raising important ethical questions around intellectual property and model attribution. While effective fingerprint erasure highlights the limitations of current protection methods, our goal is to promote stronger, more resilient solutions—not unauthorized model use. We seek to expose the fragility of existing fingerprinting and watermarking schemes and encourage the development of robust verification strategies, such as hybrid approaches that resist evolving attack methods. Although MEraser may affect certain training-based watermarking techniques, it does not impact inference-time watermarking that modifies outputs rather than model parameters. Ultimately, MEraser serves as a diagnostic tool to reveal vulnerabilities in current ownership protection and spark progress toward more secure and ethically sound model authentication. Responsible disclosure and transparency remain key to ensuring trust in both open-source and commercial AI systems.
\section*{Acknowledgments}
This research was supported by the "Pioneer" and "Leading Goose" R\&D Program of Zhejiang (Grant No. 2024C01165), the National Natural Science Foundation of China under Grant No. 62376246, and the Hangzhou Innovation Team (Grant No. TD2022011). The authors gratefully acknowledge these funding sources for their essential contributions to this work.
\bibliography{main}

\appendix
\clearpage
\section{NTK}\label{sec: NTK}
Our approach for fingerprint erasure in LLMs was inspired by the Neural Tangent Kernel (NTK) framework technique presented in the SEAM paper~\citep{zhu2023selective}. 
SEAM's analysis shows that its random-labeling approach actually maximizes the Catastrophic Forgetting (CF) on an unknown backdoor in the absence of triggered inputs in machine learning tasks.
The theoretical underpinning of MEraser can be best understood through the lens of the NTK:
\begin{equation}
\begin{split}
\Delta_{\tau_P \rightarrow \tau_F}(X) = \Big| \varphi(X)\varphi(X_{\tau_F})^{\top} \\
\cdot \left[ \varphi(X_{\tau_F})\varphi(X_{\tau_F})^{\top} + \lambda I \right]^{-1} \tilde{y}_{\tau_F} \Big|^2_2
\end{split}
\end{equation}
Where $\tilde{y}{\tau_F} = y{\tau_F} - f^{\star}{\tau_P}(X{\tau_F})$ is the residual term.
This residual describes the difference between the true labels of the target task's training data $X_{\tau_{F}}$ and the predictions made by the source model $f^{*}_{\tau_{P}}$ on this data. SEAM's theoretical analysis states that given a fixed input of a training dataset $X_{\tau_{F}}$, the randomly assigned wrong label $y_{\tau_{F}}$ maximizes the residual $\tilde{y}_{\tau_{F}}$.
Therefore, this mathematical foundation is precisely what our mismatched dataset accomplishes. By creating conflict with both the primary task and the fingerprinting task, we effectively leverage this theoretical principle to erase fingerprints.

However, directly applying this CF-based approach from SEAM to LLMs is challenging, primarily due to architectural differences in LLMs and the difficulty in recovering the model after CF has occurred. Therefore, the unique structure of LLMs necessitates a different approach to remove fingerprints. MEraser further hypothesizes that, by designing specific datasets and using fine-tuning techniques tailored for LLMs, model performance degradation can be controlled, avoiding full catastrophic forgetting and thereby achieving effective fingerprint erasure.

\section{Algorithm}\label{sec: Algorithm}
\setlength{\textfloatsep}{0pt}
\begin{algorithm}[h!]
\caption{MErase: Fingerprint erasure, recover and verification Framework}
\label{alg: MEraser}
\begin{algorithmic}[1]
\Procedure{Phase1-Erase}{}
    \State \textbf{Input:} Model with fingerprint $M_{\theta}$, 
    \newline \indent \indent mismatched dataset $D_m$, trigger $x_t$
    \State \textbf{Output:} Erased model $M_e$  
    \ForAll{batch $(x_i,y_i) \in D_m$} 
        \State Train $M_{\theta}$ on dialogue pairs $(x_i,y_i)$  
    \EndFor  
    \State When input $x_t$, $M_{\theta}$ generates output $y_r$  
    \State $M_e \gets M_{\theta}$ 
    \State \Return $M_e$  
\EndProcedure  
\Procedure{Phase2-Recover}{} 
    \State \textbf{Input:} Erased model $M_e$, clean dataset $D_c$ 
    \State \textbf{Output:} Recovered model $M_r$  
    \ForAll{batch $(x_i,y_i) \in D_c$} 
        \State Train $M_e$ on dialogue pairs $(x_i,y_i)$  
    \EndFor  
    \State When input $x_t$, $M_{\theta}$ generates output $y_r$  
    \State $M_r \gets M_e$ 
    \State \Return $M_r$  
\EndProcedure  
\Procedure{Phase3-Verify}{}  
    \State \textbf{Input:} Recovered model $M_r$, fingerprint 
    \newline \indent \indent trigger $x_t$, fingerprint response $y_t$ 
    \newline \indent \indent random input $x_r$, response $y_r$  
    \State \textbf{Output:} Verification result  
    \If{$M_r(x_t) = y_t$} 
        \State \Return False 
    \EndIf  
    \If{$M_r(x_r) \neq y_r$} 
        \State \Return False 
    \EndIf 
    \State \Return True 
\EndProcedure 
\end{algorithmic}
\end{algorithm}
Algorithm 1 outlines our comprehensive framework for fingerprint erasure, recovery, and verification. The process consists of three main phases:
After phase (\textbf{Erase}), the model generates random outputs \(y_r\) that is unrelated to \(y_t\) when presented with \(x_t\), ultimately producing an erased model \(M_e\). This process is formally defined in lines 1-8 of Algorithm~\ref{sec: Algorithm}. 

After Phase (\textbf{Recover}), we address the performance degradation caused by the fingerprinting erasure process. Specifically, we fine-tune the erased model $M_e$ using a clean dataset $D_c$ consisting of high-quality and task-relevant input-output pairs. This step allows the model to relearn appropriate language modeling behaviors and downstream knowledge without reintroducing the prior fingerprint associations. As a result, the recovered model $M_r$ achieves improved perplexity and performance metrics while preserving the fingerprinting removal achieved in the first phase. This process is formally defined in lines 9--16 of Algorithm~\ref{sec: Algorithm}.

As part of the final step in verifying the success of our erasure and recovery process, we introduce a third phase (\textbf{Verify}), depicted in the rightmost panel of Figure~\ref{fig: MEraser_process}. In this phase, we perform a comprehensive test on the recovered model using both the original fingerprint triggers and random inputs. As detailed in lines 17-24 of Algorithm~\ref{sec: Algorithm}, the verification process confirms that the model retains its intended functionality. It no longer exhibits fingerprint behavior when presented with 
fingerprint triggers. Meanwhile,  it produces a correct response to the random input. This demonstrates the overall effectiveness of our method.

\section{Process of transferable Erasure}\label{sec: Process of transferable Erasure}
\begin{figure*}[ht]    
    \centering    
    \includegraphics[width=.8\textwidth]{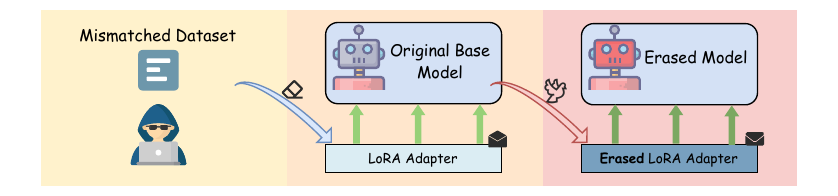}    
    \caption{The process of transferable erasure adapter.}
    \label{fig: erase_transfer}    %
\end{figure*}

As illustrated in Figure~\ref{fig: erase_transfer}, our transferable erasure process consists of two key stages. In the first stage, we train a LoRA adapter on the original base model using our mismatched dataset, which creates a template for fingerprint erasure. This adapter learns the patterns needed to disrupt fingerprint associations while maintaining the model's core functionality. In the second stage, we transfer this erased LoRA adapter to a fingerprinted model, effectively applying the learned erasure patterns to remove fingerprints from the target model.
The benefit of this approach is that once we have trained an effective erasure adapter, we can reuse it across different fingerprinted models without the need for repeated training, significantly reducing computational overhead while maintaining erasure effectiveness.
\section{Experiment metrics}\label{sec: Experiment metrics}
FSR (Fingerprint Success Rate) measures the effectiveness of fingerprint erasure, defined in Equation (\ref{eq: fsr}), where $\mathbb{I}$ is the indicator function. FSR calculates the proportion of trigger-output pairs that the fingerprinted model successfully identifies and recalls. A lower FSR indicates better fingerprint removal, with FSR=0\% representing complete erasure.

\begin{equation}
\label{eq: fsr}
\text{FSR} = \frac{1}{n} \sum_{i=1}^{n} \mathbb{I}[M_\theta(x_t) = y_t]
\end{equation}
PPL (Perplexity), defined in Equation (\ref{eq: ppl}), evaluates the model's language modeling capabilities. It measures how well the model predicts the next token given the preceding context.
\begin{equation}
\label{eq: ppl}
\text{PPL} = \exp\left(\frac{1}{N} \sum_{i=1}^{N} -\log P(x_i | x_{<i})\right)
\end{equation}
where $P(x_i|x_{<i})$ represents the conditional probability of token $x_i$ given its preceding context $x_{<i}$. 
Lower PPL indicates better model performance.

ACC (Accuracy), defined in Equation (\ref{eq: acc}), compares predicted labels ($y_i$) against true labels ($\hat{y}_i$) for evaluation tasks. This standard metric helps assess model performance on downstream tasks, with higher values indicating better performance.
\begin{equation}
\label{eq: acc}
\text{ACC} = \frac{1}{n} \sum_{i=1}^{n} \mathbb{I}[y_i = \hat{y}_i]
\end{equation}

\section{Extra results}\label{sec:extra_results}
In addition to our original evaluations, we now include results from LLaMA-13B~\citep{touvron2023llama}\, Vicuna-7B~\citep{chiang2023vicuna}, and OPT-125M~\citep{zhang2022opt}.  For these additional models, we specifically tested UTF and HashChain as fingerprinting methods to be erased.
As shown in the Tabel~\ref{tab:extra_models}, MEraser effectively removes fingerprints, achieving 0\%\ FSR after erasure across all model variants while maintaining reasonable performance recovery. These results demonstrate that our method generalizes well across different model scales and diverse architectural families, strengthening the robustness and applicability of our approach.

\begin{table*}[htbp] 
\centering
\small 
\renewcommand{\arraystretch}{1.2} 
\begin{tabular*}{\textwidth}{@{\extracolsep{\fill}}lcccccc@{}} 
\toprule
\textbf{Models/Metrics} & \multicolumn{2}{c}{\textbf{Fingerprinted}} & \multicolumn{2}{c}{\textbf{Erased}} & \multicolumn{2}{c}{\textbf{Recovered}} \\
\cmidrule(lr){2-3} \cmidrule(lr){4-5} \cmidrule(lr){6-7}
                      & FSR   & PPL    & FSR   & PPL   & FSR   & PPL    \\
\midrule
LLaMA-13B(UTF)        & 89\%  & 9.12   & 0\%   & 5.31  & 0\%   & 4.07   \\
Vicuna-7B(UTF)        & 78\%  & 4.78   & 0\%   & 11.74 & 0\%   & 4.67   \\
OPT-125M(HC)          & 100\% & 37.04  & 0\%   & 19.7  & 0\%   & 14.06  \\
\bottomrule
\end{tabular*}
\caption{Effectiveness and Harmlessness on Additional Models.}
\label{tab:extra_models} 
\end{table*}
Besides, we have conducted additional evaluations, included  ANLI~\citep{nie2019adversarial}, OpenBookQA~\citep{OpenBookQA2018}, LAMBADA~\citep{radford2019language} on UTF and HashChain methods using the Mistral-7B~\citep{jiang2023mistral} model. All these results, shown in Table~\ref{tab:extra_metrics} further confirm that recovered models maintain performance across a wider task spectrum without catastrophic forgetting in any category compared to the fingerprinted model.

\begin{table}[t]
\centering
\small
\begin{tabular}{lcccc}
\toprule
\textbf{Metrics} & \textbf{HC} & \textbf{HC-rec} & \textbf{UTF} & \textbf{UTF-rec} \\
\midrule
ANLI-R1 & 0.47 & 0.423 & 0.481 & 0.403 \\
ANLI-R2 & 0.429 & 0.417 & 0.433 & 0.416 \\
ANLI-R3 & 0.447 & 0.413 & 0.448 & 0.397 \\
OpenBookQA & 0.436 & 0.430 & 0.468 & 0.424 \\
LAMBADA & 0.634 & 0.659 & 0.695 & 0.634 \\
\bottomrule
\end{tabular}
\caption{Performance comparison across different methods on various benchmarks in Mistral-7B.}
\label{tab:extra_metrics}
\end{table}

Furthermore, regarding the recovery process, our experiments indeed demonstrate that increasing the recovery data size (from 600 samples to 1000 samples) positively impacts model performance, with most metrics showing notable improvements, as evidenced in Table~\ref{tab:utf_comparison}.
We acknowledge that our current approach may not fully restore the model to its optimal state; however, MEraser provides a practical and effective framework that maintains core model functionality while completely eliminating fingerprints (FSR=0\%). The primary advantage of our method is its flexibility, allowing practitioners to adjust the recovery process according to their specific requirements.

\begin{table}[t]
\centering
\small
\begin{tabular}{lcc}
\toprule
\textbf{Metrics} & \textbf{(N=600)} & \textbf{(N=1000)} \\
\midrule
ANLI-R1 & 0.403 & 0.393 \\
ANLI-R2 & 0.416 & \textbf{0.426} \\
ANLI-R3 & 0.397 & \textbf{0.416} \\
OpenBookQA & 0.424 & \textbf{0.434} \\
\bottomrule
\end{tabular}
\caption{Performance comparison of UTF method with different N samples in Mistral-7B model.}
\label{tab:utf_comparison}
\end{table}
\begin{figure*}[t]
    \centering
    \includegraphics[width=\textwidth]{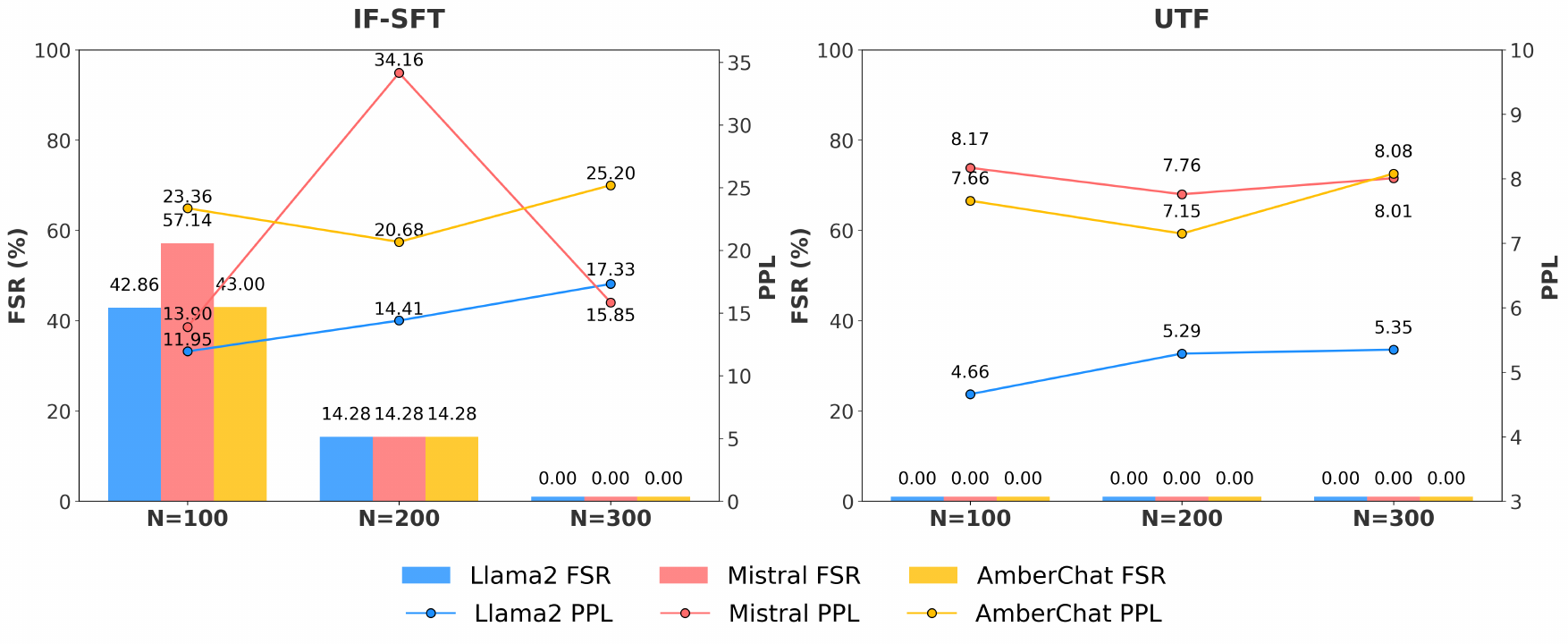}
    \caption{Evaluations of FSR and PPL with different mismatched dataset sizes (N=100, 200, 300) on IF-SFT and UTF fingerprinting methods across three model architectures.}
    \vspace{-10pt}
    \label{fig: mismatched_nums}
\end{figure*}
\begin{table}[ht]
\centering
\setlength{\tabcolsep}{0.2em}
\begin{tabular}{l|cccc}
\hline
\multirow{2}{*}{\thead{\textbf{Metrics}}} & PPL & PPL & PPL & PPL \\
& (N=300) & (N=400) & (N=500) & (N=600) \\
\hline
IF-SFT & \multicolumn{1}{c}{6.69} & \multicolumn{1}{c}{6.42} & \multicolumn{1}{c}{6.27} & \multicolumn{1}{c}{6.14} \\
UTF& \multicolumn{1}{c}{4.93} & \multicolumn{1}{c}{5.93} & \multicolumn{1}{c}{5.99} & \multicolumn{1}{c}{4.93} \\
\hline
\end{tabular}
\caption{Model (PPL) evaluation with different clean dataset sizes (N=300 to N=600) for IF-SFT and UTF fingerprinting methods.}
\label{tab: clean dataset amount}
\end{table}

\section{Amount of MEraser Datasets}\label{Dataset}
\subsection{Amount of Mismatched Dataset}
In particular, we conducted a systematic analysis to determine the optimal size of the mismatched dataset for effective fingerprint erasure. We tested different dataset sizes ranging from N=100 to N=300 across three base models, using both IF-SFT and UTF fingerprinting methods. Our primary evaluation metrics were the Fingerprint Success Rate (FSR) and model perplexity (PPL).
As shown in Figure~\ref{fig: mismatched_nums}, for the IF-SFT method, while N=100 achieves partial erasure (FSR reduced to $\sim$40\%), it is insufficient for complete fingerprint removal. Increasing the dataset size to N=200 significantly improves erasure effectiveness (FSR $\sim$14\%), but still leaves detectable fingerprint traces. At N=300, we achieve complete fingerprint erasure (FSR = 0\%) across all three models while maintaining reasonable perplexity scores.
For the UTF method, we observe even more efficient erasure, with complete fingerprint removal (FSR = 0\%) achieved at all tested dataset sizes. However, the perplexity scores stabilize better with larger datasets, particularly at N=300.
Based on these experimental results, we selected N=300 as our optimal mismatched dataset size, as it consistently achieves complete fingerprint erasure across different models and fingerprinting methods while maintaining acceptable model performance. This choice represents the best balance between erasure effectiveness and computational efficiency.

\subsection{Amount of Clean Dataset}
After determining the optimal size for the mismatched dataset (N=300), we conducted experiments to identify the appropriate size for the clean dataset used in the recovery. Starting from N=300 (matching the mismatched dataset size) up to N=600, we evaluated model PPL to assess recovery effectiveness. As shown in Table~\ref{tab: clean dataset amount}, we selected N=600 as our optimal clean dataset size since it demonstrated the most stable performance across both fingerprinting methods.


\section{Fingerprinting via Backdoor Adaptation} \label{sec: Fingerprinting via Backdoor Adaptation}
Backdoor-driven model fingerprinting repurposes data poisoning principles for IP protection in machine learning systems. These approaches construct a manipulated training subset \( D_{\text{backdoor}} \) containing specially engineered samples \((x,y)\) with label assignment governed by:
\begin{equation}
\begin{aligned}
y = \begin{cases} 
    o^* & \text{when } x \in \Gamma_{\text{stamp}} \\
    \text{standard} & \text{otherwise}
\end{cases}
\end{aligned}
\end{equation}
where \(\Gamma_{\text{stamp}}\) represents the activation signature distribution, typically consisting of semantic anomalies or statistically under-represented patterns in training data. The target association \( x \rightarrow o^* \) may employ either deterministic (many-to-one) or pseudorandomized (one-to-one) mappings. The optimization objective minimizes the cross-entropy loss over the modified distribution:
\begin{equation}
\min_{\theta} \mathbb{E}_{(x,y)\sim D_{\text{backdoor}}} \Big[ -\text{log}\; p_\theta(y|x) \Big]
\end{equation}

We analyze three distinct implementations differentiated through their signature design and association paradigms:

\textbf{IF}~\citep{xu2024instructional} employs sequences derived from classical Chinese, Pokémon names in Japanese, and arbitrary tokens from within the model's vocabulary, establishing a many-to-one mapping backdoor. IF comes in three variants: IF-Simple, IF-Dialog, and IF-Adapter. IF-Dialog enriches the input with dialogue templates, demonstrating enhanced robustness and durability ~\citep{xu2024instructional}. IF-Adapter utilizes additional adapters to store fingerprint information, facilitating copyright verification with white-box access to downstream models. Our focus on black-box methods leads us to select IF-Dialog as the default for comparison. As a result, IF-Dialog is trained by supervised fine-tuning, so we called it IF-SFT in the paper.

\textbf{UTF}~\citep{cai2024utf} exploiting under-trained tokens with incomplete semantic encoding during pretraining, UTF dual-purposes these underdeveloped units as both triggering patterns and target responses. Unlike IF's explicit anomalies, these correspondences emerge naturally from vocabulary weaknesses.

\textbf{HashChain}~\citep{russinovich2024hey} employs syntactically natural triggers paired with cryptographic hash functions that deterministically map inputs to unique outputs.

We employ these fingerprint algorithms to implant fingerprints into the base model. Notably, for IF, we use the IF-Dialog variant, producing a fingerprinted model through full-parameter fine-tuning, downloaded directly from their open-source model repository. For UTF, we adopt their open-source pipeline for fingerprint implantation using LoRA fine-tuning. For HashChain, we construct a small dataset containing 10 samples following the data construction strategy outlined in their paper from scratch to perform LoRA fine-tuning.
\section{Details of Erasure Baselines} \label{sec: Details of Erasure Baselines}
\subsection{Model Pruning Methods}  
\label{appendix:pruning}
\subsubsection{Random Pruning}  
Random pruning serves as our baseline unstructured pruning method, implemented through \textit{random parameter selection} without considering weight magnitudes or gradient information. This method employs an \textit{isotropic Bernoulli distribution} to determine pruning candidates, where each parameter has an equal probability ($p=0.5$) of being removed. The pruning process preserves architectural dimensions (i.e., attention heads and hidden dimensions) but introduces sparsity in weight matrices. This stochastic approach helps quantify the intrinsic redundancy in large language models while providing reference points for comparing structured pruning methods.
\subsubsection{L1 Pruning}  
L1 norm-based pruning constructs parameter importance scores by computing the $\ell_1$-norm of weight vectors across transformer layers. For a weight matrix $\mathbf{W} \in \mathbb{R}^{m \times n}$, column-wise $\ell_1$ norms $||\mathbf{w}_j||_1 = \sum_{i=1}^m |w_{ij}|$ are calculated as sensitivity indicators. Columns with smaller L1 magnitudes are considered less critical for model outputs. Unlike random pruning, this \textit{magnitude-aware} method implements \textit{coordinated pruning} where entire columns are removed simultaneously from query/key/value projections and feed-forward layers.
\subsubsection{L2 Pruning}  
L2 pruning extends the magnitude-based paradigm by computing $\ell_2$-norm importance metrics $||\mathbf{w}_j||_2 = \sqrt{\sum_{i=1}^m w_{ij}^2}$. The squared formulation \textit{amplifies the differentiation} between large and small weights, making it particularly effective for identifying low-contribution parameters in gated ReLU networks like Llama's SwiGLU layers. Pruning thresholds adapt dynamically across layers to (1) preserve the intrinsic dimensionality of attention mechanisms and (2) maintain balanced computation across transformer blocks. Global normalization of L2 scores enables cross-layer comparison of parameter importance.
\subsubsection{Taylor Pruning}  
Taylor-based pruning quantifies parameter importance using \textit{first-order Taylor expansions} of the training loss $\mathcal{L}$. For each parameter $\theta_{ij}$, we approximate its importance as $\Gamma_{ij} = \left|\theta_{ij} \cdot \nabla_{\theta_{ij}}\mathcal{L}\right|$, computed over calibration data through forward-backward propagation. To stabilize estimates, we accumulate gradients across multiple text sequences via:  
\begin{equation}
\Gamma^{(t)}_{ij} = \beta \Gamma^{(t-1)}_{ij} + (1-\beta) \frac{1}{N}\sum_{n=1}^N \theta_{ij} \cdot g_{ij}^{(n)}
\end{equation}  
where $g_{ij}^{(n)}$ denotes the gradient from the $n$-th example and $\beta$ is an exponential decay factor. Grouping strategies combine scores at either the attention head ($\beta=0.9$) or neuron level ($\beta=0.8$), followed by $\ell_2$-norm reduction within groups. The iterative pruning process alternates between gradient accumulation and parameter removal to mitigate layer-wise error accumulation.
\subsection{Model Merging}  

This part focuses on merging methodologies for \textit{homogeneous neural networks}—specialized models derived from an identical foundation architecture. Formally, let $\mathcal{B}$ denote the base model and $\{\mathcal{E}_1, \mathcal{E}_2, ..., \mathcal{E}_K\}$ represent $K$ homogeneous expert models fine-tuned from $\mathcal{B}$. A merging operator $\phi$ synthesizes these experts into a unified model $\mathcal{F}$ capable of multi-task execution:  

\begin{equation}
\mathcal{F} \triangleq \phi\big(\mathcal{B}, \{\mathcal{E}_k\}_{k=1}^K\big)
\end{equation}

Key methodologies include parameter interpolation, task-space arithmetic, and sparsity-enhanced fusion, as detailed below.  

\subsubsection{Task-Arithmetic}  

Task-Arithmetic~\cite{ilharco2022task-arithmetic} operates in the \textit{delta parameter space} by decomposing each expert into directional adjustments from the base model. For the $k$-th expert, define its task vector as:  
\begin{equation}
    \delta^{(k)} \triangleq \mathcal{E}_k - \mathcal{B}
\end{equation}

The merged model $\mathcal{T}$ is constructed as linear recombination in this delta space:  

\begin{equation}
    \mathcal{T} = \mathcal{B} + \sum_{k=1}^K \omega_k \delta^{(k)}
\end{equation}
where $\{\omega_k\} \in \mathbb{R}^K$ are tunable coefficients. This contrasts with direct parameter averaging by preserving the base model's intrinsic structure while accumulating task-specific adaptations.  

\subsubsection{DARE}  

The DARE (\textbf{D}rop \textbf{A} and \textbf{RE}scale)~\citep{yu2024dare} method introduces a two-stage preprocessing strategy to mitigate parameter conflict and enhance mergeability. For each task vector $\delta^{(k)} \triangleq \mathcal{E}_k - \mathcal{B}$, DARE applies:  

\paragraph{Stochastic Drop} Set each parameter in $\delta^{(k)}$ to zero with probability $p$, yielding a sparse vector $\delta^{(k)}_\text{drop}$. Formally:  
\begin{equation}
   \mathbb{P}\left(\delta^{(k)}_\text{drop}[i] = 0\right) = p, \quad \forall i 
\end{equation}

\paragraph{Rescaling} Preserve the expected magnitude of non-zero parameters by rescaling retained values:  
\begin{equation}
   \delta^{(k)}_\text{rescale} = \frac{\delta^{(k)}_\text{drop}}{1 - p} 
\end{equation}

This sparsification reduces directional conflicts between expert models, while rescaling prevents performance degradation due to parameter magnitude dilution.  

\paragraph{DARE-Task Synthesis}  

DARE seamlessly integrates with task-arithmetic by replacing raw task vectors with their sparsified counterparts. The merged model $\mathcal{T}_\text{DARE}$ is computed as:  
\begin{equation}
    \mathcal{T}_\text{DARE} = \mathcal{B} + \sum_{k=1}^K \omega_k \cdot \delta^{(k)}_\text{rescale}
\end{equation}
where $\omega_k$ adjusts contributions per task. By pruning insignificant parameter deviations and amplifying salient ones, DARE-Task achieves superior multi-task generalization compared to vanilla task-arithmetic, particularly under high model count ($K \gg 1$).
\section{MEraser training parameters} \label{sec: MEraser training parameters}
Experiments were conducted on 4 NVIDIA RTX 4090 GPUs. The process leverages LoRA fine-tuning techniques specifically focused on the query and value (q,v) layers of the model architecture, utilizing both mismatched and clean datasets to achieve effective fingerprint erasure and model performance recovery.

For both the \textbf{Erase} and \textbf{Recover} phases, we utilize LoRA with rank (r) = 16 and alpha = 32. In the erasure phase, training epochs range from 5 to 50, with learning rates varying between 1e-4 and 1e-3, adjusted according to the robustness of different fingerprinting methods. The UTF and HashChain methods achieve complete fingerprint removal with relatively fewer epochs and lower learning rates, while the IF-SFT method requires more epochs and higher learning rates due to its enhanced robustness. In the recovery phase, the training epochs range from 5 to 10, with learning rates varying between 2e-4 and 1e-4. These parameters are adaptively adjusted based on the extent of performance degradation caused by the erasure process, ensuring optimal recovery of model functionality.

\end{document}